\documentstyle[prl,aps,preprint]{revtex}
\begin{document}

\clubpenalty=10000
\widowpenalty=10000
\brokenpenalty=10000
\interdisplaylinepenalty=5000
\predisplaypenalty=10000
\postdisplaypenalty=100
\tolerance=1000

\tighten

\title{SOFFER'S INEQUALITY \thanks
{This work is supported in part by funds provided by the U.S.
Department of Energy (D.O.E.) under cooperative
agreement \#DF-FC02-94ER40818 and \#DE-FG02-92ER40702.}}

\author {Gary R. Goldstein}

\address{Department of Physics \\
Tufts University \\
Medford, Massachusetts 02155}

\author{R. L. Jaffe and Xiangdong Ji}

\address{Center for Theoretical Physics \\
Laboratory for Nuclear Science \\
and Department of Physics \\
Massachusetts Institute of Technology \\
Cambridge, Massachusetts 02139 \\
{~}}

\date{MIT-CTP-2402 \hfill HEP-PH /9501297
           \hfill Submitted to: {\it Physical Review D1}
           \hfill January 1995}

\maketitle

\begin{abstract}
Various issues surrounding a recently proposed inequality
among twist-two quark distributions in the nucleon are
discussed. We provide a rigorous derivation of
the inequality in QCD, including radiative corrections
and scale dependence. We also give a more heuristic, but
more physical derivation, from which we show that
a similar inequality does not exist among twist-three
quark distributions. We demonstrate that the inequality
does not constrain the nucleon's tensor charge.
Finally we explore physical mechanisms
for saturating the inequality, arguing it is unlikely
to occur in Nature.

\end{abstract}

\pacs{xxxxxx}

\section{Introduction}

In a recent letter~\cite{SOFFER}, Soffer has proposed a new inequality among
the nucleon's twist-two quark distributions, $f_1, g_1$, and
$h_1$~\cite{JAFJI1,JAFJI2},
\begin{equation}
f_1 + g_1\ge 2|h_1| \; .
 \label{inequality}
\end{equation}
$f_1$ is the well-known spin average quark distribution which measures
the probability to find a quark in a nucleon independent of its spin
orientation.  $g_1$ measures the polarization asymmetry in a
longitudinally polarized nucleon --- the probability to find a quark
polarized along the nucleon's spin minus the probability to find a
quark polarized against the nucleon's spin. $h_1$, which is less
familiar, measures the polarization asymmetry in a transversely
polarized nucleon.  $f_1$ and $g_1$ have been measured in many deep
inelastic scattering experiments.  $h_1$ decouples from lepton
scattering and has not yet been measured.  Proposals to measure $h_1$
at HERA and RHIC have generated efforts to characterize $h_1$, hence
the interest in this inequality~\cite{PARTICLEWORLD,HERMES}.

Soffer derives Eq.~(\ref{inequality}) by analogy between quark-nucleon
scattering and nucleon-nucleon scattering, where helicity amplitudes
analagous to $f_1$, $g_1$, and $h_1$ obey inequalities derived many
years ago~\cite{OLDSOFFER}.  There are potential problems with this
analogy.  The intermediate states in quark-hadron scattering, which
are treated as on-shell physical states in Soffer's derivation, are,
in fact, colored and gauge dependent.  The distribution functions
$f_1$, $g_1$, and $h_1$ are, in fact, {\it integrals} of quark-hadron
forward scattering amplitudes over transverse momentum with cutoffs at
$k_\perp\approx\sqrt{Q^2}$. In QCD, the definitions of quark
distributions such as $f_1$, $g_1$, and $h_1$ are scale and
renormalization scheme dependent.  Any relations among them must be
accompanied by a precise description of the procedure with which they
are extracted from experimental data. In contrast, the well-known
inequalities and positivity constraints among distribution functions
such as $f_1\ge\vert g_1\vert$ are general properties of lepton-hadron
scattering, derived without reference to quarks, color and QCD.

In this Paper we consider Soffer's inequality in the context of QCD\@.
We find that Eq.~(\ref{inequality}) can be derived in a ``parton model
approximation'' to QCD, but that radiative corrections modify
Eq.~(\ref{inequality}) in a significant way.  Each term in
Eq.~(\ref{inequality}) is multiplied by a power series in
$\alpha_s(Q^2)/\pi$.  So the inequality as presented by Soffer is of
limited practical use --- it is strictly valid only at asymptotic
$Q^2$ where $\alpha_s\to 0$ and the distribution functions vanish for
all $x>0$.  Thus the inequality has a similar status in QCD as the
Callan--Gross relation~\cite{CALLANGROSS} --- a parton model result
which is invalidated by QCD radiative corrections.  One should
remember, however, that the Callan--Gross relation is a very useful,
although approximate tool in deep-inelastic phenomenology.  A one-loop
calculation of the radiative corrections to Eq.~(\ref{inequality}),
which we have not attempted, would yield an improved result which
would be useful at experimentally accessible $Q^2$.

In \S II we study Soffer's inequality from the consideration of {\it
current}--hadron scattering amplitudes.  This treatment has the same
level of rigor as the derivation of standard deep-inelastic
inequalities such as $f_1\ge\vert g_1\vert$, and demonstrates the
presence of radiative corrections in QCD\@.  In \S III we present a
second derivation closer in spirit to Soffer's earlier analysis to
nucleon-nucleon scattering.  This derivation is heuristic.  In
particular, it ignores QCD radiative corrections.  However, it enables
us to make contact with standard operator definitions of the
distributions $f_1$, $g_1$, and $h_1$. It is then straightforward to
generalize the analysis to twist-three (corrections of ${\cal
O}({1/\sqrt{Q^2}})$).  In his paper Soffer suggested that there would
be a twist-three generalization of his
inequality~\cite{SOFFER}. Although there is a natural correspondence
between the three twist-two distributions, $f_1$ $g_1$ and $h_1$, on
the one hand, and the three twist-three distributions, $e$, $g_T$, and
$h_L$, on the other~\cite{JAFJI1}, we find that there is {\it no\/}
such inequality at twist three.  Also in his paper, Soffer claims that
the inequality places a constraint on the nucleon's ``tensor charge,''
the lowest moment of $h_1$.  Using the formalism of \S III we show
that this result is invalidated by the presence of antiquarks in the
nucleon wavefunction and that there is no way to define the notion of
a ``valence quark'' to give a useful result.

Soffer noted that his inequality appeared to be saturated for single
quarks in simple quark models such as the non-relativistic quark model
and the bag model~\cite{SOFFER,JAFJI1}.  In \S IV we demonstrate that
this feature is not preserved by even the simplest quark model
wavefunctions.  For example, the inequality is saturated for
down-quarks in the quark model proton, but not for up-quarks.  Also,
saturation is not preserved by evolution.  We comment on the
possibility of using saturation ({\it e.g.}~for down-quarks in the
proton) as ``boundary data''~\cite{PP,JR}.

\section{Derivation of the Inequality from Current-Hadron Amplitudes}

It is useful to review the textbook derivation of the inequalities or
``positivity constraints'' on the familiar structure functions of deep
inelastic lepton scattering, $f_1$, $f_2$, $g_1$, and
$g_2$~\cite{TEXT}.  They follow from demanding that cross sections for
forward, vector current-hadron scattering are positive definite.
These cross sections are proportional to
\begin{eqnarray}
W(\epsilon) & = & {1\over 4\pi}\sum_X \left( 2\pi\right)^4
\delta^4(P+q-P_X)\Vert \langle X\vert J\cdot\epsilon\vert P, S\rangle \Vert^2
           \; , \nonumber \\
 & = & \epsilon^{\mu *}W_{\mu\nu}\left(q,P,S\right)\epsilon^\nu  \; ,
\label{positive}
\end{eqnarray}
which is manifestly positive definite for any $\epsilon$.  $P^\mu$ and $S^\mu$
are the momentum and spin of the target ($P^2 = -S^2 = M^2$,
$P\cdot S = 0$), and $\epsilon^\mu$ is the polarization
vector of the (virtual) photon.
$J_\mu$ is the electromagnetic current operator, which
in QCD would be  $\sum_a e_a{\bar{\psi}}^a\gamma_\mu\psi^a$,
where $a$ is a flavor label. For simplicity we
consider a single quark flavor with unit charge. Hence the
relations we derive will be valid for each flavor separately.
$W_{\mu\nu}$ is the usual current-current correlation
function of deep inelastic scattering,
\begin{eqnarray}
W_{\mu\nu} (q,P,S) & = & {1\over 4\pi}\int d^4\xi e^{iq\cdot\xi}\langle P,
S\vert
\left[ J_\mu (\xi), J_\nu(0)\right]\vert P, S\rangle \ \ ,\nonumber \\
& = & -g_{\mu\nu}f_1(q^2,\nu)
+{1\over\nu}P_\mu P_\nu f_2(q^2,\nu)+{\rm gauge\  terms}\nonumber \\
& & \quad {} +
	{i\over\nu}\epsilon_{\mu\nu\rho\lambda}q^\rho S^\lambda g_1(q^2,\nu)
+{i\over\nu^2}\epsilon_{\mu\nu\rho\lambda}q^\rho \left(\nu
S^\lambda-q\cdot S P^\lambda\right)g_2(q^2,\nu)
\label{wmunu}
\end{eqnarray}
where $q^2 < 0$, and $\nu=P\cdot q>0$. Substituting this expansion
back into Eq.~(\ref{positive}) and taking the Bjorken scaling limit
yields $f_1 + g_1\ge 0$ or $f_1 - g_1\ge 0$ for transverse photons and
definite nucleon helicity states, hence $f_1\ge\vert g_1\vert$.

The current $\bar{\psi}^a\gamma_\mu\psi^a$ creates and annihilates
antiquarks as well as quarks so the structure functions all receive
both quark and antiquark contributions. In the Bjorken limit,
$\lim_{Bj}$ ($Q^2=-q^2,\nu\to\infty, x\equiv - q^2/{2\nu}$ fixed) of
QCD, $f_1$ and $g_1$ reduce to quark {\it distribution functions\/}
which we label with the flavor $a$ or $\bar a$ of quark or antiquark,
\begin{eqnarray}
\lim_{Bj} f_1(q^2,\nu) & = & f_1^a(x,\ln Q^2) + f_1^{\bar a}(x,\ln Q^2)
           \;  ,
\nonumber\\
\lim_{Bj} g_1(q^2,\nu) & = & g_1^a(x,\ln Q^2) + g_1^{\bar a}(x,\ln Q^2)
           \;  ,
\label{scaling}
\end{eqnarray}
and the positivity constraints will apply to such combinations.
The physical meaning of the inequality can be seen from
the fact that the combination $f_1+g_1$ in parton model is
simply the probability to find a quark or antiquark with spin
parallel to the target nucleon,
\begin{equation}
\lim_{Bj} [f_1(q^2,\nu) + g_1(q^2,\nu)] =
q^{\uparrow a}(x,\ln Q^2) +
\bar q^{\uparrow a}(x,\ln Q^2) \;  .
\label{quarkspin}
\end{equation}
and conversely for $f_1-g_1$.  The $\ln Q^2$ dependence comes from the
evolution of the distributions under scale transformation.  Note that
these distributions have been defined in terms of deep-inelastic
vector-current structure functions.  Quark distributions are in
general process-dependent and relations among quarks distributions
extracted from different experiments can be calculated in QCD
perturbation theory~\cite{ALTARELLI}.

Of course the quark and antiquark distributions $f_1^a\ , g_1^a$ and
$f_1^{\bar{a}}\ , g_1^{\bar{a}}$ are separately constrained.  We must
understand how this comes about in order to obtain the strongest
possible bounds that include the transverse structure function, $h_1$.
We would like to replace $J_\mu$ by a current which couples only to
quarks.  The {\it chiral\/} currents $J^\pm_\mu ={1\over 2}(V_\mu\pm
A_\mu)$, which are given by ${1\over
2}\bar\psi\gamma_\mu(1\pm\gamma_5)\psi$ in QCD, are candidates.
$J^-_\mu$, for example, couples to left-handed quarks and right-handed
antiquarks.  If we choose the polarization vector, $\epsilon^\mu$,
judiciously, we can select left-handed quanta, thereby decoupling the
antiquarks.  To be specific, we choose the momentum $\vec q$ to be in
the positive $\hat e_3$ direction, $q^\mu = (q^0,0,0,q^3)$, and $\vec
P$ to be in the $- \hat e_3$ direction.  If we employ the V--A
current, negative helicity for the target nucleon, and $\epsilon_-^\mu
= {1\over\sqrt{2}}(0,1,-i,0)$, then the current selects left-handed
quarks and right-handed antiquarks in the left-handed target:
$q^{\downarrow a} +\bar q^{\uparrow a}$.  The right-handed antiquarks
decouple from the product $J^-\cdot\epsilon_-$ because they have
${\cal J}_z = -{1\over 2}$ and cannot absorb $\Delta{\cal J}_z = -1$.
It is quite easy to see that $f_1^a + g_1^a\ge 0$ results.  Analogous
choices yield constraints on $f_1^a-g_1^a$ and on antiquark
distributions.

The derivation we have just outlined would be quite complicated for
non-asymptotic $q^2$ and $\nu$.  The introduction of chiral currents
and polarized targets requires all the machinery developed for
neutrino scattering from polarized targets~\cite{JINEUTRINO}. Such an
analysis would lead to a very general constraint, valid independent of
QCD and the Bjorken limit.  However, it is only {\it useful\/} in the
Bjorken limit where only the familiar twist-two invariant structure
functions $f_1$ or $g_1$ survive.  The same remark will apply in the
case of Soffer's inequality to which we now turn.

The quark currents $\bar\psi\gamma_\mu\psi$ and
$\bar\psi\gamma_{\mu}\gamma_5\psi$ preserve quark chirality.  So does
the leading term in the product of two such currents at short
distances.  The distribution function $h_1$, in contrast, couples
quarks of opposite chirality \cite{JAFJI1} and therefore does not
appear in any of these relations .  This suggests that constraints
involving $h_1$ might be obtained by considering the {\it
interference\/} between the V$-$A current and a current of opposite
chirality.  This is in fact the case. So, in addition to the V$-$A
current, $J^-_\mu$, we introduce a hypothetical current, ${\cal J}$,
which is composed of scalar and pseudoscalar currents, along with
tensor and pseudo-tensor currents
\begin{equation}
{\cal J}\equiv (S + P - T^{+-} - T_5^{+-})/2\sqrt{2} \; .
\label{funny}
\end{equation}
This ungainly choice has been engineered to select out the
distribution functions of interest.  Unlike the vector and axial
currents which are defined by symmetries, these currents cannot be
defined independent of quarks and QCD.  For example, different
constraints on distribution functions would be obtained from
$S=\bar\psi\psi$ or $S=\bar\psi\psi\bar\psi\psi$.  We define the
currents as follows: $S(\xi) = Z_S\bar\psi(\xi)\psi(\xi)$, $P(\xi) =
Z_P\bar\psi(\xi)\gamma_5\psi(\xi)$, $T^{\mu\nu}(\xi) =
Z_T\bar\psi(\xi){1\over 2}[\gamma^\mu,\gamma^\nu]\psi(\xi)$, and
$T_5^{\mu\nu}(\xi) = Z_{T_5} \bar\psi(\xi){1\over
2}[\gamma^\mu,\gamma^\nu]\gamma^5\psi(\xi)$.  Because these currents
are not constrained by Ward-identities, they are non-trivially
renormalized in QCD.  As a consequence in addition to the ambiguities
already mentioned, they are regularization and renormalization scheme
dependent. However, for any choice of scheme, the derivation of the
inequality remains the same, and, of course, the physical implications
of the inequality are scheme independent.
%
%One in principle can exploit this degree of freedom
%to obtain a most stringent constraint.
%
For simplicity, however, we choose dimensional regularization and
(modified) minimal subtraction.  The renormalization scale in currents
is set at the virtual-boson mass, $Q^2$.  The tensor and pseudo-tensor
currents combine with the scalar and pseudo-scalar currents to project
the ``good'' light-cone components of the right-handed chiral fermions
(as will be discussed in the next section) from the field $\psi$. When
positive helicity is chosen for the nucleon, the right-handed quark
field will remain, rather than the left-handed anti-quark.

The desired inequality follows from consideration of a judiciously chosen
fictitious ``cross section.'' Consider the quantity,
\begin{eqnarray}
{\cal W}(q,P) & = & {1\over{4\pi}}\sum_X \left( 2\pi\right)^4
\delta^4(P+q-P_X)\Vert \langle X\vert J_-\cdot\epsilon_-\ \vert P-\rangle \pm
\langle X\vert {\cal J}\vert P+\rangle
\Vert^2
            \; , \nonumber \\
 & = & {1\over 4\pi}\int d^4\xi e^{iq\cdot\xi}\left[\langle P, -\vert
\left[ J_-^\dagger(\xi)
\cdot\epsilon_-^*, J_-(0)\cdot\epsilon_-\right]\vert P, -\rangle
 + \langle P, +\vert
\left[ {\cal J}^\dagger(\xi), {\cal J}(0)\right]\vert P, +\rangle\right]
            \nonumber \\
 & & \quad {}  \pm  {1\over 2\pi}{\rm Re}
\int d^4\xi e^{iq\cdot\xi}\langle P, +\vert
\left[ {\cal J}^\dagger(\xi), J(0)
\cdot\epsilon_-\right]\vert P, -\rangle  \; ,
\label{bigpositive}
\end{eqnarray}
which is manifestly positive. ${\cal W}$ involves three terms.
Referring back to Eq.~(\ref{wmunu}) it is clear that the
$J_-^\dagger\cdot\epsilon_-^*\otimes J_-\cdot\epsilon_-$ term will
reduce to $f_1^a + g_1^a$ in the Bjorken limit.  Likewise, it is clear
from general considerations that the ${\cal J^\dagger\otimes J}$ term
will also involve $f_1^a$ and $g_1^a$ in the Bjorken limit.  However,
since ${\cal J^\dagger\otimes J}$ suffers different radiative
corrections than $J_-^\dagger\cdot\epsilon_-^*\otimes
J_-\cdot\epsilon_-$, $f_1$ and $g_1$ will be multiplied by a series in
$\alpha_s(Q^2)/\pi$.  The interference term, ${\cal J^\dagger}\otimes
J\cdot\epsilon_-$, is chiral-odd and can only involve $h^a_1$ in the
Bjorken limit.  Combined with the other two terms, we obtain
\begin{equation}
\lim_{Bj}{\cal W} =  R_f(\alpha_s(Q^2))
  f^a_1(x,\ln Q^2)+R_g(\alpha_s(Q^2))g^a_1(x,\ln Q^2)\pm
 2R_h(\alpha_s(Q^2))h^a_1(x,\ln Q^2)  \;.
\label{identity}
\end{equation}
Here the $R_f$ and $R_g$ factors take into account the radiative
corrections mentioned above.  The $R_h$ factor arises because the
definition of $h_1$ is process dependent.  If we chose to {\it
define\/} $h_1$ through our fictitious process then $R_h=1$ {\it by
definition\/}.  On the other hand, if $h_1$ is defined through a
physical process such as Drell-Yan $\mu$-pair production with
transversely polarized beams~\cite{JAFJI1}, then $R_h$ is $1 + {\cal
O}(\alpha_s)$.  Another subtlety in this calculation is that the
vector-scalar interference terms have the (nucleon) helicity structure
$\langle P\pm\vert\dots\vert P\mp\rangle$, which does not correspond
to an expectation value in a state of definite spin.  However the
helicity structure required can be extracted by combining expectation
values in states with $\vec S = \hat e_1$ and $\vec S = \hat e_2$.
Radiative corrections aside, the result is straightforwardly obtained
by calculating the current correlation functions at tree-level in the
Bjorken limit, and using the standard definitions of the distribution
functions $f_1^a$, $g_1^a$, and $h_1^a$~\cite{JAFJI1}.

Since ${\cal W}$ is manifestly positive, eq (\ref{identity}) is the
desired inequality.  Of course ${\cal W}$ is positive for all $q^2$
and $\nu$. So (\ref{bigpositive}) implies a constraint among the many
invariant structure functions that occur in the decomposition of
${\cal W}$ at sub-asymptotic $q^2$ and $\nu$.  There is no point,
however, in displaying this inequality explicitly, since nearly all
the novel structure functions, such as those involved in the invariant
decomposition of $T_{\mu\nu}\otimes J^\pm_\lambda$, are not directly
measurable.

This derivation shows that Soffer's inequality holds independently for
each quark and antiquark flavor.  Also, it is clear that careful
attention must be given to the specific ``process'', in which the
quark distributions can be defined unambiguously. The ``natural''
choice would be to define $f_1$ and $g_1$ in vector-current
deep-inelastic scattering, and $h_1$ in polarized Drell-Yan. It is
clear that Soffer's identity is a parton model approximation (no
radiative corrections) to a more useful identity which can be obtained
by computing the factors $R_f$, $R_g$, and $R_h$ at least through
(lowest non-trivial) order $\alpha_s/\pi$.

Armed with this rigorous, if rather unphysical, derivation, we turn to
examine the inequality from the more familiar viewpoint of the quark
parton model and its coordinate space equivalent, the light-cone
expansion.

\section{Derivation of the Inequality from Quark Hadron Amplitudes}

We begin with a simple, heuristic ``parton model'' derivation of the
inequality postponing any complexity.  Next we introduce the bilocal
light-cone correlation functions which allow us to give a more
convincing derivation and study twist-three distribution functions.
Only QCD radiative corrections will be left out at this stage.  The
derivation of the previous section shows how their effects can be
included.

In the most elementary parton model, deep inelastic processes are
summarized by the ``handbag'' diagram of Fig.~1a.  At the bottom of
this diagram is the imaginary part of a quark-nucleon scattering
amplitude.\footnote{The propagator on the quark leg is not truncated.}
We focus on this amplitude.  Since the quark (nucleon) begins and ends
with the same momentum, $k$ ($P$), the amplitude describes {\it
forward\/} scattering.  Since the quark is initially {\it removed\/}
from the nucleon and then {\it replaced\/}, the diagram actually
corresponds to a {\it u-channel} discontinuity of forward
quark-nucleon scattering, as shown in Fig.~1b. We label the
$u$-channel discontinuities ${\cal A}_{Hh,H'h'}$, where $H$ and $H'$
are the initial and final nucleon helicities and $h$ and $h'$ are the
{\it outgoing\/} and {\it incoming\/} quark helicities
respectively. For spin-1/2 quarks and nucleons parity and
time-reversal invariance reduce the number of independent helicity
amplitudes to three.  Three convenient choices shown in Fig.~2, are
${\cal A}_{++,++}$, ${\cal A}_{+-,+-}$, and ${\cal A}_{++,--}$
respectively. Amplitudes that fail to satisfy conservation of angular
momentum along the collision axis, $H+h'=H'+h$, vanish.  Other
helicity amplitudes are either related to these by parity, ${\cal
A}_{Hh,H'h'}={\cal A}_{-H-h,-H'-h'},$ or time reversal, ${\cal
A}_{Hh,H'h'}={\cal A}_{H'h',Hh}.$ It is easy to show that the three
twist-two structure functions, $f_1$, $g_1$ and $h_1$ are (suitably
normalized) linear combinations of ${\cal A}_{++,++}$, ${\cal
A}_{+-,+-}$, and ${\cal A}_{++,--}$~\cite{JAFJI2}.
$f_1 = {\cal A}_{++,++} + {\cal A}_{+-,+-}$,
$g_1 = {\cal A}_{++,++} - {\cal A}_{+-,+-}$, and
$h_1 = {\cal A}_{++,--}.
$

To obtain the Soffer's inequality it is necessary to consider the
quark-hadron amplitudes which are related to the \{${\cal A}$\} by
unitarity.  Define four amplitudes $a_{Hh}$ by
\begin{equation}
a_{Hh}(X) = \langle X|\phi_h|P H\rangle \; ,
\label{amplitudes}
\end{equation}
where $\phi$ is the quark field, and $X$ is an arbitrary final state.
Unitarity requires that the \{${\cal A}$\} are proportional to
products of the form $\sum_{X}a_{H'h'}^*(X)a_{Hh}(X)$, so
\begin{eqnarray}
f_1  &\propto &\sum_{X}a_{++}(X)^*a_{++}(X)+ a_{+-}(X)^*a_{+-}(X)
	\ , \nonumber \\
g_1  &\propto &\sum_{X}a_{++}(X)^*a_{++}(X)- a_{+-}(X)^*a_{+-}(X)
	\ , \nonumber \\
h_1  &\propto &\sum_{X}a_{++}(X)^*a_{--}(X)\ ,
\label{unitarity}
\end{eqnarray}
The desired inequality follows from the observation that
\begin{equation}
\sum_X\Vert a_{++}(X) \pm a_{--}(X)\Vert^2\ge 0  \; ,
\label{amplitude}
\end{equation}
and that ${\cal A}_{++,++}={\cal A}_{--,--}$ and
${\cal A}_{++,--}={\cal A}_{--,++}$ by parity.

Our first step in improving this admittedly heuristic derivation is to
clarify the relationship between the helicity amplitudes $\{{\cal
A}\}$ and $\{a\}$ and the operator expressions which define the
distribution functions $f_1$, $g_1$, and $h_1$ in QCD\@.  First we
will derive Eqs.~(\ref{unitarity}) from standard definitions of $f_1$,
$g_1$, and $h_1$.  Then it will be straightforward to show that the
inequality does not generalize to twist-three.  Also it will be clear
that the tensor charge is not constrained by Eq.~(\ref{inequality}).

In QCD parton distributions are defined by the light-cone Fourier
transformation of forward matrix elements of operator products.  The
quark distributions of interest to us are related to matrix elements of
bilinear quark operators,
\begin{eqnarray}
\int {d\lambda\over{4\pi}}e^{i\lambda x}\langle PS|\bar\psi(0)\gamma_\mu
\psi(\lambda n)|PS\rangle
& = & f_1(x)p_\mu + M^2f_4(x)n_\mu
           \label{f1}\\
\int {d\lambda\over{4\pi}}e^{i\lambda x}\langle
PS|\bar\psi(0)\gamma_\mu\gamma_5
\psi(\lambda n)|PS\rangle
& = & g_1(x)p_\mu S\cdot n +[g_1(x)+g_2(x)]S_{\perp\mu}\nonumber\\
& & \quad {} + M^2g_3(x)n\cdot Sn_\mu
           \label{g1}\\
\int {d\lambda\over{4\pi}}e^{i\lambda x}\langle PS|\bar\psi(0)
\psi(\lambda n)|PS\rangle
& = & Me(x)
           \label{e1}\\
\int {d\lambda\over{4\pi}}e^{i\lambda x}\langle
PS|\bar\psi(0)\sigma_{\mu\nu}i\gamma_5\psi(\lambda n)|PS\rangle
& = &  h_1(x)(S_{\perp\mu}p_\nu-S_{\perp\nu}p_\mu)/M\nonumber\\
& & \quad {} + [h_2(x)+h_1(x)/2]M(p_\mu n_\nu-p_\nu n_\mu)S\cdot n\nonumber\\
& & \quad {} + h_3(x)M(S_{\perp\mu}n_\nu- S_{\perp\nu}n_\mu)
           \label{h1}
\end{eqnarray}
where $n$ and $p$ are null vectors of mass dimension $-1$ and $1$,
respectively ($n^2=p^2=0$, $n^+=p^-=0$, $n\cdot p =1$).  $P$ and $S$
may be decomposed in terms of $n$ and $p$, $P=p+{M^2\over 2}n$, $S_\mu
= S\cdot np_\mu +S\cdot p n_\mu +S_{\perp\mu}$.  For a target moving
in the ${\bf \hat e_3}$ direction, $ p = {1\over\sqrt 2}({\Lambda}, 0
, 0, {\Lambda}), n = {1\over\sqrt 2}({1\over {\Lambda}}, 0 , 0,
-{1\over {\Lambda}})$.  In Eqs.~(\ref{f1}--\ref{h1}) $\psi$ is the
four-component Dirac field for the quark.  The flavor label on $\psi$
and the corresponding distribution functions has been suppressed.

 Eqs.~(\ref{f1}--\ref{h1}) are written in $n\cdot A=0$ gauge.  In any
other gauge a Wilson link would be required between $\psi$ and
$\bar\psi$ to maintain gauge invariance.  Gluon radiative corrections,
which generate a renormalization point dependence for these operators
and an associated $q^2$ dependence for the distribution functions,
have been suppressed in Eqs.~(\ref{f1}--\ref{h1}).

The leading twist contributions to Eqs.~(\ref{f1}--\ref{h1}) are the
distributions functions $f_1$, $g_1$, and $h_1$ respectively.  They
may be projected out by contracting the equations with $n^\mu$,
$n^\mu$, and $n^\mu S^{\perp\nu}$ respectively.  In every case the
projection operator $ {\cal
P}^+\equiv\gamma^0(\gamma^0+\gamma^3)/2=(1+\alpha_3)/2 $ emerges from
the Dirac algebra.  ${\cal P}^+$ projects the four component Dirac
spinor $\psi$ onto the two dimensional subspace of ``good'' light-cone
components which are canonically independent fields~\cite{KOGUTSOPER}.
Likewise, $ {\cal
P}^-\equiv\gamma^0(\gamma^0-\gamma^3)/2=(1-\alpha_3)/2 $ projects on
the two dimensional subspace of ``bad'' light-cone components which
are interaction dependent fields and should not enter at leading
twist~\cite{JAFJI2}. Much of our analysis is simplified by choosing a
representation for the Dirac matrices tailored to the
light-cone~\cite{KOGUTSOPER},
\begin{equation}
\gamma^0 = \rho_1\sigma_3,\
\gamma^1 = i \sigma_1,\
\gamma^2 = i \sigma_2,\
\gamma^3 = -i \rho_2\sigma_3,\
\gamma_5 = i \gamma^0\gamma^1\gamma^2\gamma^3 =\rho_3\sigma_3\ .
\label{gammas}
\end{equation}
Where $\{\rho_k\}$ and $\{\sigma_k\}$ are $2\times 2$ Pauli matrices.
This is to be contrasted to the familiar Dirac-Pauli representation, $
\{\rho_3,
i\rho_2 \sigma_1,
i\rho_2 \sigma_2,
i\rho_2 \sigma_3,
\rho_1\}
$ which is convenient for many other purposes.  In the light-cone
representation ${\cal P}^\pm$, ${\gamma_5}$ and $\vec\sigma\cdot \hat
e_3$ are all diagonal,
\begin{equation}
{\cal P}^+ =\left(\matrix{ {\bf 1} & {\bf 0} \cr {\bf 0} &  {\bf 0}}\right),\
{\cal P}^- =\left(\matrix{ {\bf 0} & {\bf 0} \cr {\bf 0} &  {\bf 1}}\right),\
\gamma_5 =\left(\matrix{ \sigma_3 & {\bf 0} \cr {\bf 0} &
-\sigma_3}\right),\
\vec\sigma\cdot\hat e_3=\left(\matrix{\sigma_3 & {\bf 0} \cr {\bf 0} &
\sigma_3}\right)\ .
\label{dirac}
\end{equation}
where $ {\bf 1}$ and ${\bf 0}$ are the $2\times 2$ identity and null
matrices respectively.  In this basis ${\cal P}^+$ and ${\cal P}^-$
project onto the upper and lower two components of the Dirac spinor
respectively,
\begin{equation}
\phi\equiv {\cal P}^+\psi=\left(\matrix{\phi_+\cr \phi_-\cr}\right),\
\chi\equiv {\cal P}^-\psi=\left(\matrix{\chi_+\cr \chi_-\cr}\right)  \; .
\label{phichi}
\end{equation}
$\phi_\pm$ are the ``good'' light-cone components of the quark field,
which are independent canonical variables in the light-cone
formulation.  $\chi_\pm$ are the ``bad'' light-cone components which
may be regarded as composite fields built from quarks (the ``good''
light-cone components) and transverse gluons.  The $\pm$ labels on
$\phi$ and $\chi$ refer to the eigenvalue of $\sigma_3$ which is
proportional to {\it helicity\/}, $\vec s\cdot\hat P$, for quarks
moving in the $\hat e_3$ direction,
($\vec\sigma\cdot\hat P\phi_\pm=\pm\phi_\pm,
\vec\sigma\cdot\hat P\chi_\pm=\pm\chi_\pm$),
not to {\it chirality\/}.  From Eqs.~(\ref{dirac}) and (\ref{phichi})
it is clear that helicity and chirality are the same for $\phi$, but
opposite for $\chi$. This is easy to understand when one recognizes
that the bad light-cone components are actually composites of the
canonically independent operators $\phi_\pm$ and $\vec A_\perp$.  The
positive helicity component of $\chi$ ($\chi_+$) involves a transverse
gluon (with positive helicity) and a good light-cone component of the
quark field, $\phi_-$ (with negative helicity and therefore negative
chirality).

It is now straightforward to project $f_1$, $g_1$, and $h_1$ out of
Eqs.~(\ref{f1} --\ref{h1}) and rewrite the result in terms of $\phi_\pm$,
\begin{eqnarray}
f_1(x) & = & \int{d\lambda\over{4\pi{\cal P}}}e^{i\lambda x}
\langle P+\vert\phi_+^\dagger(0)\phi_+(\lambda n)
+\phi_-^\dagger(0)\phi_-(\lambda n) \vert P+\rangle
           \nonumber \; , \\
g_1(x) & = & \int{d\lambda\over{4\pi{\cal P}}}e^{i\lambda x}
\langle P+\vert\phi_+^\dagger(0)\phi_+(\lambda n)
-\phi_-^\dagger(0)\phi_-(\lambda n) \vert P+\rangle
           \nonumber \; , \\
h_1(x) & = & \int{d\lambda\over{8\pi{\cal P}}}e^{i\lambda x}
\{\langle P+\vert\phi_+^\dagger(0)\phi_-(\lambda n)\vert P-\rangle
+\langle P-\vert\phi_-^\dagger(0)\phi_+(\lambda n) \vert P+\rangle\}
           \; ,
\label{lc-distributions}
\end{eqnarray}
If we insert a complete set of intermediate states between
$\phi^\dagger$ and $\phi$, translate the fields and carry out the
$\lambda$ integration, we obtain,
\begin{eqnarray}
f_1(x) & = & {1\over{2{\cal P}}}\sum_X\delta(x-1+n\cdot P_X)
\{\vert a_{++}(X)\vert^2+\vert a_{--}(X)\vert^2\}
           \nonumber \;  , \\
g_1(x) & = & {1\over{2{\cal P}}}\sum_X\delta(x-1+n\cdot P_X)
\{\vert a_{++}(X)\vert^2-\vert a_{--}(X)\vert^2\}
           \nonumber \;  , \\
h_1(x) & = & {1\over{2{\cal P}}}\sum_X\delta(x-1+n\cdot P_X)
a_{++}(X)^*a_{--}(X)
           \;  .
\label{lc-distributions2}
\end{eqnarray}
% *** I changed the label on the last eqn so that reference is made to
% the previous eqn in the following paragraph. ***
This reproduces Eq.~(\ref{unitarity}) and shows that the ``generic''
quark fields which appear there should be identified with the chiral
components of the ``good'' light-cone components of the quark field.
This derivation illustrates the questionable procedure required to
obtain Soffer's inequality using traditional parton-model/light-cone
methods: the states in $\vert X\rangle$ are colored; and the bilocal
operators in Eq.~(\ref{lc-distributions}) do not actually exist since
each term in their Taylor expansion about $\lambda=0$ is renormalized
differently by radiative corrections.  However the result is correct
(modulo the important radiative corrections discussed in \S II) and
the derivation is considerably more ``physical'' than the more
rigorous one presented in the previous section.

The light-cone formalism defined in this section allows us to examine
the possible extension of Soffer's identity to the twist-three
distributions, $e$, $g_T$, and $h_L$.  $e$ is defined in
Eq.~(\ref{e1}), and the others are defined by, $g_T=g_1+g_2$ and
$h_L={1\over 2}h_1+h_2$.  Examination of Eqs.~(\ref{f1}--\ref{h1})
shows that $e(q^2,\nu)$ is spin-independent and chiral-odd.  $h_L$ and
$g_T$ are spin-dependent and chiral-odd and chiral-even respectively.
$h_L$ is associated with longitudinally polarized targets and $g_T$
with transversely polarized targets.  In summary, the spin attributes
of $\{e,h_L,g_T\}$ correspond to $\{f_1,g_1,h_1\}$
respectively.\footnote{For further discussion of $\{e,h_L,g_T\}$,
see~\cite{JAFJI2}.}  The astute reader will note that this
correspondence appears to be inconsistent with the chirality
assignments of the distribution functions.  For example, $f_1$ is spin
average, and therefore diagonal in helicity --- $f_1\propto
\phi^\dagger_+\phi_++\phi^\dagger_-\phi_-$.
Clearly $f_1$ preserves quark chirality
--- {\it i.e.\/}~it is chiral-even.  $e$ on the other hand is claimed
to be chiral-odd, even though it, like $f_1$, averages over helicity.
The resolution of this apparent contradiction comes from the
classification of $e$ with respect to the light-cone projection
operators ${\cal P}_\pm$.  It is easy to see that $e\propto
\chi^\dagger_+\phi_++\chi^\dagger_-\phi_- + {\rm h.c.}$.
A glance at the chirality
assignments of $\chi_\pm$ confirms that $e$ flips chirality -- {\it
i.e.\/}~it is chiral-odd.  An analogous analysis applies to $h_L$ and
$g_T$.

It should now be clear that an identity analagous to
Eq.~(\ref{inequality}) {\it cannot\/} be obtained at twist-three.  The
reason is that an object of the form
$\langle\Psi\vert\chi^\dagger_\pm\phi_\pm\vert\Psi\rangle$ could only
arise by starting with positive definite structure such as
$\Vert\langle X\vert\chi_\pm\vert\Psi\rangle+\langle
X\vert\phi_\pm\vert\Psi\rangle\Vert^2$.  This object would generate
twist-three distributions in the interference, but twist-two, and more
problematically, {\it twist-four\/} distributions such as
$\langle\Psi\vert\chi^\dagger_ \pm\chi_\pm\vert\Psi\rangle$ would be
unavoidable.  The conclusion then is that any positivity constraint
involving the twist-three distributions $e$, $g_T$, and $h_L$ would
inextricably include twist-four distributions which are very difficult
to measure.  Hence Soffer's speculation is incorrect~\cite{SOFFER}.

Finally we consider the relationship imposed on the lowest moment of
$h_1$ by the inequality, Eq.~(\ref{inequality}).  The nucleon's {\it
tensor charge\/}, $\delta q^a(Q^2)$ is defined by analogy to the axial
charge, $\Delta q^a$~\cite{JAFJI1},
\begin{eqnarray}
S^k\delta q^a(Q^2) & \equiv & {1\over 2}\langle PS\vert\bar
q\sigma^{0k}i\gamma_5{\lambda^a\over 2} q\mid_{Q^2}\vert PS\rangle =S^k
\int^1_{-1}dx \lbrack h^a_1(x,\ln Q^2)-h^{\bar a}_1(x,\ln Q^2)\rbrack
           \; , \nonumber \\
S^k\Delta q^a & \equiv & {1\over 2}\langle PS\vert\bar
q\gamma^k \gamma_5 {\lambda^a\over 2} q\vert PS\rangle =S^k \int^1_{-1}dx
\lbrack g^a_1(x,\ln Q^2)+g^{\bar a}_1(x,\ln Q^2)\rbrack
           \; .
\label{tensor}
\end{eqnarray}
In contrast to the nucleon's {\it axial charge} which figures in
beta-decay, the tensor charge does not appear in weak matrix elements
and has not been measured.  Note that $\delta q^a$ is renormalization
point dependent, whereas $\Delta q^a$ is not (because the axial
current in QCD is conserved apart from quark mass terms).  Note also
that $h^{\bar a} (g^{\bar a})$ enters Eq.~(\ref{tensor}) with a minus
(plus) sign reflecting that the operator $\bar
q\sigma^{\mu\nu}\gamma_5 q\ \ (\bar q\gamma^\mu\gamma_5 q)$ is odd
(even) under charge conjugation.  There is no way to combine
Eqs.~(\ref{tensor}) with the inequalities $f^a_1+g^a_1\ge 2\vert
h^a_1\vert$, and $f^{\bar a}_1+g^{\bar a}_1\ge 2\vert h^{\bar
a}_1\vert$ to obtain any useful information about $\delta q^a$ without
further assumptions.  Soffer~\cite{SOFFER} suggests that his
inequality applies to the {\it valence\/} quark distributions in the
nucleon, however the only circumstances in which we find a useful
bound is if we assume that the nucleon contains no antiquarks at all
($f^{\bar a}_1 = g^{\bar a}_1 = h^{\bar a}_1=0$), which is known to be
false.

\section{Saturation of Soffer's Inequality}

There are some special circumstances for which Soffer's inequality is
saturated, {\it i.e.\/}\break
$2\vert h^a_1(x)\vert =
f^a_1(x)+g^a_1(x)$. It is useful to consider such cases in order to
develop some intuition about the distribution of spin within the
nucleon and to speculate on how saturation may be used to estimate
$h_1(x)$ in regions of experimental interest. The most trivial case is
a model in which all the spin and flavor information of the proton is
carried by a single quark, either in a non-relativistic quark model
(NRQM) or the bag model. In the NRQM, if two quarks are always in a
spin and flavor scalar configuration, then the third quark will yield
$h^a_1(x)=f^a_1(x)=g^a_1(x)$ --- a consequence of the rather trivial
Dirac structure of non-relativistic spinors.  The bag model is less
trivial due to the lower component $p$-wave contribution. Nonetheless,
the saturation remains valid. In more realistic case of an SU(6) wave
function, the saturation only holds for the d-quark, as we will
demonstrate below.

The possibility of saturation is related to a possible symmetry
between the amplitudes $a_{++}(X)$ and $a_{--}(X)$ defined in
Eq.~(\ref{amplitudes}).  In particular, if $a_{++}(X)=a_{--}(X)$ {\it
for all} states $X$ contributing to the sums which define $f_1$, {\it
etc.\/} in Eq.~(\ref{unitarity}), then the inequality is saturated
with the $+$-sign for the absolute value.  To relate $a_{++}$ to
$a_{--}$ consider the unitary operator, $U$ defined as the product of
parity, $\Pi$, and a rotation by 180$^\circ$ about an axis
perpendicular to $\hat P$, $U\equiv\Pi R_2(\pi)$.  Here we have chosen
$\hat P$ to define the $\hat e_3$-axis and rotated (by $\pi$) about
the $\hat e_2$-axis. It is easy to see that $U$ transforms $\vert
P+\rangle$ into $\vert P-\rangle$ up to a phase.  Likewise, $U$
transforms $\phi_+$ into $\phi_-$ up to a phase [Note that
$\Pi^\dagger\psi(0)\Pi=\gamma^0\psi(0)=\rho_1\sigma_3\psi(0)$ and
$R_2(\pi)^\dagger\psi(0)R_2(\pi)=-i\rho_2\sigma_2\psi(0)$, so
$U^\dagger\psi(0)U=-i\rho_3\sigma_1\psi(0)$.] Applying this
transformation to $a_{++}(X)$ we obtain,
\begin{equation}
a_{++}(X)= {\rm phase}\times a_{--}(UX).
\label{ua}
\end{equation}
So the saturation of the identity resolves down to the question of
whether $X$ is an eigenstate of the operator $U$.  In simple valence
quark models, the state $X$ consists merely of the two spectator
quarks left behind when the operator $\phi_+$ annihilates one quark in
the target state $\vert P +\rangle$.

First consider, for definiteness, the {\it down\/} quark distribution
in a simple constituent quark model of the proton.  The two spectator
$u$--quarks must be in a ${\cal J}=1$ state on account of Fermi
statistics.  Thus the angular momentum structure of the wavefunction
is,
\begin{equation}
\vert \hat P = \hat e_3\ + \rangle = \sqrt{2\over 3}\vert \{uu\}^{{\cal J}=1,
{\cal J}_3=1}d^\downarrow\rangle
	- \sqrt{1\over 3}\vert \{uu\}^{{\cal J}=1,{\cal
J}_3=0}d^\uparrow\rangle  \;.
\label{wavefunction}
\end{equation}
Only the second term contributes to $a_{++}$, leaving the spectator
state $\vert X\rangle =\break - \sqrt{1\over 3}\vert\{uu\}^{{\cal
J}=1,{\cal J}_3=0}\rangle$, which clearly is an eigenstate of $U$.  A
careful accounting of all the phases yields
\begin{equation}
a_{++}(X)= - \eta_P\eta^3_q a_{--}(X)
\label{phases}
\end{equation}
where $\eta_P$ ($\eta_q$) is the intrinsic parity of the nucleon (quark)
and the negative sign arises from the conventional Condon and Shortley
phases in the Clebsch-Gordon series.
Since the relative parity of the quark and nucleon is positive, the factor
$ - \eta_P\eta^3$ is minus one, and the inequality is saturated with the
absolute value of $h^d_1$. The structure functions are in the ratios
$(f^d_1:g^d_1:h^d_1) = (1:-{1\over 3}:-{1\over 3})$. However, due to
the effects of $p$-wave, the saturation does not occur for
$d$ quark in the bag model.

For the {\it up\/} quark distribution in the proton the situation is
different. The spectator $u$ and $d$ quarks are in a mixed spin state,
${\cal J} = 1$ and $0$. Annihilating a $u$-quark with positive
helicity in Eq.~(\ref {wavefunction}) leaves the spectator state
$\vert X\rangle = \sqrt{1\over 10} \vert\{ud\}^{{\cal J}=1,{\cal
J}_3=0} +3\{ud\}^{{\cal J}=0}\rangle$. Annihilating a $u$-quark with
{\it negative\/}
helicity in the proton with {\it negative\/} helicity leaves the state
$\vert X\rangle = \sqrt{1\over 10} \vert\{ud\}^{{\cal J}=1,{\cal
J}_3=0} -3\{ud\}^{{\cal J}=0}\rangle$. The relative sign change for
the ${\cal J}=0$ and ${\cal J}=1$ parts means that $a_{++}(X)$ is not
a simple multiple of $a_{--}(X)$ --- there is no analog of Eq.~(\ref
{phases}) and hence, no saturation. In fact $(f^u_1:g^u_1:h^u_1) =
(2:{4\over 3}:{4\over 3})$ for the NRQM.

We see that the saturation of Soffer's inequality for the $d$-quark
follows from the particularly simple spin structure of the nucleon in
quark models.  It is easy to construct a more elaborate model in which
even that saturation fails.  For example, suppose we introduce a
component into the nucleon wavefunction in which the spectators are
coupled to total angular momentum ${\cal J} = 0$, say,
$\vert\{uug\}^{{\cal J}=0}d^\uparrow\rangle$, where $g$ is a gluon.
Then the state $X$ is a superposition of components, one with ${\cal
J}={\cal J}_3=0$ and the other with ${\cal J}=1, {\cal J}_3=0$.  These
two components transform with opposite sign under $U$ and thereby ruin
Eq.~(\ref {ua}).

Finally we consider the relationship between QCD evolution and
saturation of the inequality.  Since $f_1$, $g_1$ and $h_1$ evolve
differently with $Q^2$, saturation is incompatible with evolution.  We
can understand this in light of the discussion of the previous
paragraph --- evolution mixes gluons (and $q\bar q$ pairs) into the
nucleon wavefunction destroying the simple structure responsible for
saturation.  Quark model relationships, like saturation of the
inequality for d-quarks in the proton, should be interpreted as
``boundary data'' for QCD evolution~\cite{PP,JR}, valid at some low
scale $\mu_0^2$.  The implications for experiments carried out at much
larger scales must be obtained by evolution from $\mu_0^2$ to the
experimental scale, $Q^2$.  In the case of saturation, some remnant of
a prediction for the down quark contribution to $h_1$ in the proton
might be obtained when good data on the $d$-quark contributions to
$f_1$ and $g_1$ become available.

\section {Acknowledgements}

We thank Jacques Soffer for discussions and a prepublication copy of
Ref.~\cite{SOFFER}.

\begin{figure}
\def\figurename{\hspace*{-1.5em}FIG.}
\parindent=0in
\bigskip\bigskip
\caption{a). The hand-bag diagram for deep-inelastic
scattering. b) Quark-nucleon scattering amplitudes in $s$ and $u$
channels. The momentum and helicity labels are shown
explicitly.}
\label{fig1}
\bigskip
\caption{Three independent helicity amplitudes in $u$-channel.
\hspace{14.5em}}
\label{fig2}

\end{figure}

\end{document}